\documentclass[rnote]{aa} 
\usepackage{natbib}
\usepackage{graphicx}
\usepackage{txfonts}

\usepackage{amssymb}
\usepackage[notref,notcite]{} 

\def\cm3{~{\rm cm^{-3}}}

\def\lesssim{\buildrel < \over {_{\sim}}}
\def\gtrsim{\buildrel > \over {_{\sim}}}

\begin{document}

\title{Spectra of accelerated particles at supernova shocks in the presence of neutral hydrogen: the case of Tycho}
\titlerunning{Spectra of accelerated particles in Tycho SNR in the presence of neutral hydrogen}
\authorrunning{G. Morlino \& P. Blasi}

\author{G. Morlino\inst{1,2}\fnmsep\thanks{email: giovanni.morlino@gssi.infn.it}, 
            P. Blasi\inst{2,1}\fnmsep\thanks{email: blasi@arcetri.astro.it}
            }
\institute{$^1$Gran Sasso Science Institute (INFN), viale F. Crispi 7, 67100 L'Aquila, Italy\\
              $^2$INAF/Osservatorio Astrofisico di Arcetri, Largo E. Fermi, 5, 50125, Firenze, Italy
               }

\date{Received 16 Nov. 2015 ; accepted 28 Feb. 2016}

\abstract
{The presence of neutral hydrogen in the shock proximity changes the structure of the shock and affects the spectra of particles accelerated through the first-order Fermi mechanism. This phenomenon has profound implications for the interpretation of the multifrequency spectra of radiation from supernova remnants.}
{Neutrals that undergo charge exchange with hot ions downstream of the shock may result in fast neutrals moving towards the upstream gas, where they can suffer additional charge exchange or ionisation reactions, thereby depositing energy and momentum upstream. Here we discuss the implications of this neutral return flux, which was already predicted in our previous work on neutral mediated supernova shocks, and show how the spectra of accelerated particles turn out to be appreciably steeper than $p^{-4}$, thereby affecting the gamma ray spectra from supernova remnants in general and from Tycho specifically.}
{The theory that describes non-linear diffusive shock acceleration in the presence of neutral hydrogen has been developed in recent years. Here we use a semi-analytical theory developed in previous work and specialise our predictions to the case of the Tycho supernova shock, where there is evidence from gamma ray observations that the spectrum of the parent cosmic rays is steeper than expected from the traditional theory of diffusive shock acceleration.} 
{We show that, if the fraction of neutral hydrogen in the vicinity of the Tycho supernova shock is, as suggested by observations, $\sim 70-90\%$, then spectra of accelerated protons steeper than $p^{-4}$ may be a natural consequence of charge exchange reactions and the associated neutral return flux. The spectral shape is affected by this phenomenon for particles with energies below $\sim 100-1000$ GeV, for which the diffusion length is less than or at most comparable to the pathlength of charge exchange and ionisation upstream of the shock.  
}
{}

\maketitle

\section{Introduction}
\label{sec:intro}

Based on energetic grounds, supernova remnants (SNRs) represent the most likely sources of Galactic cosmic rays (CRs) \cite[see][for a recent review]{blasirev}. The kinetic energy of the expanding shell is expected to be partially converted to CR energy by means of diffusive shock acceleration (DSA), as predicted by the non-linear theory of DSA (NLDSA) \cite[see][and references therein]{blasirev} and as seen in numerical simulations of the formation of non-relativistic collisionless shocks \cite[]{dam1,dam2,dam3}. While the energy conversion efficiency does not appear to be particularly problematic, the ability of standard SNRs to accelerate particles to energies that resemble the knee, $\sim (1-5)\times 10^{6}$ GeV, is all but trivial for assessing at the present time. Magnetic field amplification plays a crucial role in this discussion: reaching high energies requires amplification to occur in the the so-called CR driven 
regime \cite[]{bell2004}, but in this case the conditions in which the supernova explosion occurs seem to determine how high the maximum energy can be. For instance, \cite{schure1,schure2} and \cite{cardillo} point out that only core collapse supernova (SN) explosions taking place in the wind of their pre-supernova red giant parent star could possibly achieve maximum energy in the PeV region (within a few tens of years after the explosion), while standard SNe type Ia, exploding in the interstellar medium (ISM), can only reach maximum energy around a few hundred TeV. Realistic estimates of the maximum energy may be affected by damping processes. 

In general, the spectrum of accelerated particles is predicted to have a concave shape as a result of the dynamical reaction of accelerated particles on the shock, although for acceleration efficiency of $\sim 10\%$, the concavity is rather mild. As a result, the spectrum remains quite close to being a power law $\sim p^{-4}$, at least for strong shocks. This prediction has been noted to be in some friction with the spectra of gamma rays from SNRs as detected by Fermi-LAT \cite[]{damiano2011}. This friction has become even more evident in the most recent Fermi-LAT catalogue of SNRs \cite[]{FermiCat}, where the slopes of the gamma ray spectra of the SNRs detected by Fermi-LAT have been characterized in terms of different observables. It is clear from these observations that the gamma-ray spectra lie mainly above a slope $\sim 2$ (in energy) with a few exceptions (most likely these are the cases where the gamma ray emission is predominantly leptonic in origin). 

Spectra steeper than the canonical $p^{-4}$ are not easy to accommodate in the context of either the test particle theory of diffusive shock acceleration (DSA) or in terms of its non-linear extension (NLDSA). It has been proposed \cite[]{damiano2012} that in cases where magnetic field amplification is very efficient, the velocity of the scattering centres may be substantially increased with respect to the standard Alfv\'en speed \citep[however, for a general derivation of the velocity of scattering centres see][]{bykov} and that the spectra of accelerated particles may be steeper as a consequence of the larger compression factor of the velocity of scattering centres between upstream and downstream of the shock. Although this solution is viable, it is rather model dependent and relies on poorly known aspects of the micro-physics of the acceleration process. Moreover, at present, no such effect has ever been seen in simulations of the acceleration process. On the other hand, including the velocity of the scattering centres, though in a phenomenological way, has been shown to be able to reproduce the multifrequency observations of the Tycho SNR surprisingly well \cite[]{morlinotycho}. We notice, however, that  the same $\gamma$-ray emission has also been explained in terms of a complex morphology of the emission region around the shock \cite[]{berezhkotycho}.

The presence of neutral hydrogen in the shock region has also been shown to lead to steeper spectra of accelerated particles \cite[]{paper1}, at least at energies low enough that the diffusion length upstream of the shock is smaller than the path length for charge exchange and ionisation reactions. This condition is most easily fulfilled for shock speeds below 3000-4000 km/s. This effect, originally predicted in the context of a test particle theory of particle acceleration, was later included in a non-linear theory of diffusive shock acceleration \cite[]{paper3}.

In this paper we aim at applying this theory to derive the gamma ray emission from the Tycho SNR for different levels of neutral hydrogen density in the shock region. As discussed above, Tycho is one of best known cases in which spectra of accelerated particles appreciably steeper than $E^{-2}$ have been inferred from gamma ray observations \cite[]{fermi,veritas}. Some recent gamma-ray observations by Veritas \cite{} seem to suggest a spectral steepening (or even a cutoff) at gamma-ray energies of $\sim TeV$. A rather steep spectrum of accelerated particles in Tycho has also been inferred from some radio observations: the measurements of \cite{tychoradio} suggest an electron spectrum $\propto E^{-(2.2\div 2.3)}$ in the $1-10$ GeV range, although a somewhat harder integrated spectrum, $\propto E^{-2.04}$, was previously obtained by \cite{Katz-Stone00}.

The Tycho SNR originated in a type Ia SN, and it is probably expanding in the warm neutral phase of the ISM. In fact the detection of Balmer-dominated shocks with emission seen in H$\alpha$ from the eastern and northern limbs reveals the presence of a large amount of neutral hydrogen, close to $\sim 80\%$ \citep{Kamper78, Ghavamian00}. Recently, \cite{Williams13} have found evidence of an overall gradient in the ambient density surrounding Tycho, with densities $\sim 3-10$ times higher in the northeast region than in the southwest region. This large density gradient is qualitatively consistent with the variations in the proper motion of the shock observed in radio \citep{Reynoso97} and X-ray studies \citep{Katsuda10}. For an estimated distance between 2 and 3 kpc, the measurements of the proper motion return an average shock speed $\sim 4000-5000$ km s$^{-1}$, while in the northeastern region the shock speed drops to $\sim 2000-3000$ km s$^{-1}$. These considerations suggest that the morphology of the emission region in the Tycho SNR might be rather complex, despite the type Ia nature of this remnant. For instance, if an appreciable fraction of the gamma ray emission were to come from the northeastern limb, the presence of neutral hydrogen would substantially affect the spectra of accelerated particles. In this paper we discuss this possibility in a more quantitative way. 

The paper is structured as follows. In \S \ref{sec:neutral} we summarize the physical aspects of the neutral induced shock modification and we outline the theory of NLDSA when both the dynamical reaction of accelerated particles and neutrals is taken into account. In \S \ref{sec:results} we illustrate the main implications of this theoretical framework for the gamma ray emission from the Tycho SNR. We conclude in \S \ref{sec:conclude}.

\section{Particle acceleration in the presence of neutral hydrogen}
\label{sec:neutral}

The structure of a shock in the presence of neutral hydrogen has been investigated by \cite{paper1}, where the authors also discuss the acceleration of test particles at a shock modified by the neutral gas. The essential physical element of the theory is that neutral atoms can cross a collisionless shock without feeling the discontinuity developed by thermodynamical quantities, while the ionized gas is slowed down and heated up at the shock. On the other hand, neutrals are coupled to the background ions through charge exchange and ionisation reactions: neutrals try to form a collisional shock on spatial scales that are close to the path length of charge exchange (CE) and ionisation, typically $\sim 10^{16}-10^{17}$ cm.

A neutral atom crossing the shock front can suffer a CE reaction with a hot (slower) ion downstream of the shock. There is a finite probability that this reaction gives rise to a fast neutral atom moving towards the shock front (from downstream to upstream). Such an atom can recross the shock front and eventually deposit energy and momentum upstream, a phenomenon that we refer to as the {\it \emph{neutral return flux}} (NRF). This phenomenon, first found by \cite{paper1}, is responsible for the formation of a neutral induced precursor, where the gas temperature is higher and the gas is slower than the same quantities at upstream infinity. The neutral return flux leads to a weakening of the collisionless shock, so that test particles accelerated there diffusively are forced to have a steeper spectrum, provided their energy is such that their path length does not exceed the charge exchange/ionization path length \cite[]{paper1}. The cross-sections of these processes are such that this effect is especially important for shock speed $\lesssim 4000$ km/s. 

The non-linear generalisation of the theory, in which both neutral atoms and accelerated particles are allowed to modify the shock structure, was presented by \cite{paper3}. The basic aspects of the shock phenomenology discussed above remain qualitatively valid, but the situation becomes appreciably more complex: first, the dynamical reaction of accelerated particles is responsible for the usual CR induced precursor, whose size is close to the diffusion length of the highest energy particles, $\sim D(p_{max})/v_{\rm sh}$, where $D(p_{max})$ is the diffusion coefficient at the maximum momentum $p_{max}$, calculated in the amplified magnetic field. This, by itself, is a very non-linear quantity. The non-linear dynamical reaction of CRs on a collisionless shock has been discussed by many authors \citep[see][for a recent review and a list of references]{blasirev}. The main consequences of these non-linear effects are the appearance of concave spectra and magnetic field amplification as due to streaming instability. 

Neutral atoms upstream of the shock react to the slowing down of the ionized gas through effective CE in the CR induced precursor, before entering the downstream region, where the NRF originates as discussed above.
Neutral hydrogen interacts with ions through CE and ionization and with electrons through ionization alone. The hydrogen distribution function, $f_{N}(\vec v,z)$, can be described using the stationary Boltzmann equation
\begin{equation} \label{eq:vlasov}
v_z \frac{\partial f_{N}(\vec v, z)}{\partial z} = \beta_{N} f_{i}(\vec v, z)  -
        \left[ \beta_{i} + \beta_e \right] f_{N}(\vec v, z) \,,
\end{equation}
where $z$ is the distance from the shock (which is located at the origin), $v_z$ is the velocity component along the $z$ axis and the electron and proton distribution functions, $f_i(\vec v,z)$ and $f_e(\vec v,z)$, are assumed to be Maxwellian at each position. The collision terms, $\beta_k f_l$, describe the interaction (due to CE and/or ionization) between the species $k$ and $l$. The interaction rate $\beta_k$ is formally written as
\begin{equation} \label{eq:beta_k}
\beta_k (\vec v,z) = \int d^{3} w \, v_{rel} \, \sigma(\vec v_{rel})
                  f_{k}(\vec w,z) \,,
\end{equation}
where $v_{rel} = |\vec v- \vec w|$ and $\sigma$ is the cross section for the relevant interaction process. More precisely, $\beta_N$ is the rate of CE of an ion that becomes a neutral, $\beta_i$ is the rate of CE plus ionization of a neutral due to collisions with protons, while $\beta_e$ is the ionization rate of neutrals due to collisions with electrons. A full description of the cross sections used in the calculations can be found in \cite[]{paper2}.

The dynamics of the background plasma is affected by the presence of accelerated particles and by CE and ionization of neutrals. Protons and electrons in the plasma are assumed to share the same local density, $\rho_i(z)=\rho_e(z)$, but not necessarily the same temperature, i.e., $T_{i}(z)$ may be different from $T_{e}(z)$. The equations describing the conservation of mass, momentum, and energy taking the interactions of the plasma fluid with CRs into account are

\begin{equation} \label{eq:rh1}
 \frac{\partial}{\partial z} \left[\rho_i u_{i} + \mu_N  \right]=0 \,,
\end{equation}
\begin{equation} \label{eq:rh2} 
 \frac{\partial}{\partial z} \left[ \rho_i u_{i}^{2} + P_{g} + P_{c} + P_{w} 
        + P_{N}  \right]=0 \,,
\end{equation}
\begin{equation} \label{eq:rh3}
 \frac{\partial}{\partial z} \left[ \frac{1}{2} \rho_i u_{i}^{3} + 
  \frac{\gamma_{g} P_{g} u_{i}}{\gamma_{g}-1} + F_w + F_{N} \right]
  = -u_i \frac{\partial P_c}{\partial z} + \Gamma P_w \,.
\end{equation}
Here $\mu_N = m_H \int d^{3} v v_{\parallel} f_{N}$, $P_N = m_H \int d^{3} v v_{\parallel}^{2} f_{N}$ and $F_N = m_H/2 \int d^{3} v v_{\parallel} (v_{\parallel}^{2} + v_{\perp}^{2}) f_{N}$ are the fluxes of mass, momentum, and energy of neutrals along the $z$ direction, respectively. They can be computed easily once the neutral distribution function is known. In Eq. \ref{eq:rh3}, $P_{w}$ is the pressure in the form of waves, and $\Gamma$ is the rate of damping of such waves. In the same equation, $P_c$ is the CR pressure computed from the CR distribution function:
\begin{equation}
 P_c(z) = \frac{4 \pi}{3} \int dp \, p^3 v(p) f_c(z,p) \,.
\end{equation}
The latter is known once the distribution function of accelerated particles $f_{c}$ is calculated from the transport equation:
\begin{equation} \label{eq:trasp_CR}
 \frac{\partial}{\partial z} \left[ D(z,p) \frac{\partial f_c}{\partial z}
 \right]
 - u \frac{\partial f_c}{\partial z} 
 + \frac{1}{3} \frac{d u}{d z} \, p \frac{\partial f_c}{\partial p} 
 + Q(z,p) = 0  \,.
\end{equation}
The $z$-axis is oriented from upstream infinity $(z=-\infty)$ to downstream infinity $(z=+\infty)$ with the shock located at $z=0$. We assume that the injection occurs only at the shock position and is monoenergetic at $p=p_{\rm inj}$, so the injection term can be written as $Q(z,p)= Q_0(p) \delta(z)$ where $Q_0(p)= ({\eta_{\rm inj} n_1}/{4 \pi p_{\rm inj}^2}) \delta(p-p_{\rm inj})$. Here $n_1$ is the number density of ions immediately upstream of the subshock, and $\eta_{\rm inj}$ is the fraction of particles that is going to take part in the acceleration process. Following \cite{BGV05}, $\eta_{\rm inj}$ can be related to the subshock compression factor as
\begin{equation}
 \eta_{\rm inj} = 4/ \left(3 \sqrt{\pi} \right) \left(R_{\rm sub}-1 \right)
             \xi_{\rm inj}^3 e^{-\xi_{\rm inj}^2} \, .
\end{equation}
Here $\xi_{\rm inj}$ is defined by the relation $p_{\rm inj} = \xi_{\rm inj} p_{th,2}$ , where $p_{th,2}$ is the momentum of the thermal particles downstream, and $\xi_{\rm inj}$ parametrizes the poorly known microphysics of the injection process and is taken as a free parameter with a typical value between 2 and 4. We notice that a similar value has recently been inferred from the numerical simulations of \cite{dam4}.
   
\begin{figure}
\centering 
  \includegraphics[width=\linewidth]{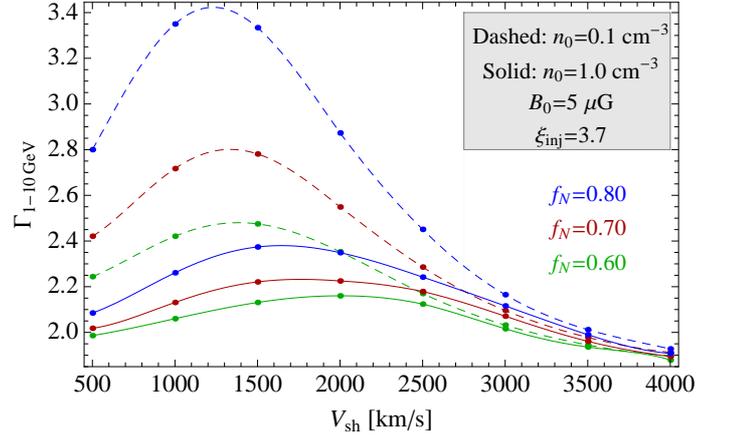}
  \caption{Slope of the gamma-ray spectrum produced by pp collisions in the energy band $1-10$ GeV for $n_{0}=1~\rm cm^{-3}$ (solid lines) and $n_{0}=0.1~\rm cm^{-3}$ (dashed lines). From top to bottom, the lines for each case refer to neutral fraction $f_{N}=0.8,~0.7$ and 0.6.}
  \label{fig:slope}%
\end{figure}

The calculations presented by \cite{paper3} consist of an iterative procedure that allows one to solve simultaneously the hydrodynamical equations for the gas made of ions and neutral atoms (Eqs. \ref{eq:rh1}-\ref{eq:rh3}), the Boltzmann equation for neutral atoms coupled to ions through CE and ionisation (Eq. \ref{eq:vlasov}), the transport equation for accelerated particles (Eq. \ref{eq:trasp_CR}), and the transport equation for the waves generated by accelerated particles through streaming instability. We refer to the paper by \cite{paper3} for the mathematical details of the theory, while here we concentrate upon the choice of the parameters' values suitable for the Tycho SNR that may allow us to find a reasonable fit to the spectrum of gamma rays from this remnant, after taking pion production and decay into account. 

\section{Results}
\label{sec:results}

The iterative calculation outlined in the previous section has been carried out for the case of the Tycho SNR, with a few important differences with respect to the method illustrated by \cite{paper3}. First, we use the amplification scheme proposed by \cite{bell2004} for the turbulent magnetic field, so that the strength of amplified field is assigned as
\begin{equation}
\delta B = n_{CR}(>E) E \frac{v_{\rm sh}}{c},
\label{eq:deltaB}
\end{equation}
where $n_{CR}(>E)=\int_{p_{\max}}^{\infty} 4\pi p^{2} f_{c}(p) dp$ is the integral spectrum of particles accelerated at the shock, calculated using the solution of the transport equation at the shock. This prescription returns a magnetic field upstream of the shock that is independent of the pre-existing magnetic field. A different recipe for saturation of the instability was proposed by \cite{riq}. Such a prescription typically leads to an amplified magnetic field $\delta B$ that is weaker that the one returned by Eq. \ref{eq:deltaB}. Moreover $\delta B$ retains a weak dependence on the value of the pre-existing magnetic field $B_{0}$. The growth of non-resonant modes is expected to proceed very fast in the upstream region of the shock and eventually slow down, while the scale of the unstable modes approaches the Larmor radius of the particles in the current generating the instability. Given the fast rise of the instability and the non-Alfvenic nature of the excited modes, in our calculations we neglected damping (i.e. $\Gamma=0$ in Eq. \ref{eq:rh3}). The detailed analysis of the growth of the non-resonant modes in a partially ionized medium carried out by \cite{Reville07} suggests that for reasonable values of the parameters the instability is not quenched. We assume that the diffusion coefficient is Bohm-like, with $D(p)=\frac{1}{3} v(p) \frac{pc}{e \delta B}$. 

Second, since our main purpose is to illustrate the effect of neutrals in the shock region rather than providing a detailed fit to observations, in the calculations below we assume that electrons and ions are in thermal equilibrium upstream of the shock ($T_{i}=T_{e}$), while downstream of the shock we assume that $\beta_e\equiv T_{e}/T_{i} \approx 0.1 $. This choice appears to be in good qualitative agreement with observations when the shock speed is $\gtrsim 1000$ km s$^{-1}$ (see e.g. \cite{Ghavamian13} for a review and \cite{Morlino14}) and also with results from PICS simulations \citep{ParkJ15}. We also notice that values of $\beta_e$ that are below 0.1 do not change our results, while if $\beta_e$ approaches unity, the ionization rate of neutrals due to electrons increases and the effect of the NRF is reduced.

The solution of the equations described in the previous section provides us with the complete shock structure, including the distribution function and hence the NRF. As pointed out above, the slope of the spectrum of accelerated particles in the presence of NRF and dynamical reaction of accelerated particles becomes a function of energy. At energies that are low enough for the diffusion length to be smaller than the CE/ionization path length upstream of the shock, the spectrum  is expected to become appreciably steeper than $E^{-2}$. Figure \ref{fig:slope} shows the slope of the gamma-ray spectrum produced by accelerated hadrons through pion decay in the gamma ray energy range $1-10$ GeV as a function of the shock velocity for $n_{0}=1~\rm cm^{-3}$ (solid lines) and $n_{0}=0.1~\rm cm^{-3}$ (dashed lines). For each of these two cases, from bottom to top the lines refer to neutral fractions $f_{N}=0.6,~0.7,$ and 0.8, in line with the values inferred by \cite{Kamper78} and \cite{Ghavamian00}. In these calculations we assumed an injection parameter $\xi_{\rm inj}=3.7$, which returns acceleration efficiencies of $\sim 10\%$. (To be more precise the efficiency increases from few percent for $v_{\rm sh}=500$ km s$^{-1}$ to $\sim 20\%$ for $v_{\rm sh}= 4000$ km s$^{-1}$.) 

Increasing the density of the background gas leads to a shorter path length for CE reactions, so that the effect of the return flux is limited to lower energies, which is reflected in the lower curves (solid lines) in Fig. \ref{fig:slope}. At any given energy, lowering the density leads to steeper spectra. At any given shock speed and total density, the effect of the NRF increases, as expected, for a larger neutral fraction $f_{N}$, so that the spectrum of accelerated particles becomes steeper. At low shock speeds, the cross-section for CE reactions is large enough to warrant a substantial NRF, so the spectrum of accelerated particles is predicted to be steeper than $E^{-2}$. At a shock speed above 3000 km/s, the cross section for CE starts dropping, so the strength of the effect decreases. The standard slope of the spectrum of accelerated particles is eventually recovered for shock speed $v_{\rm sh}\gtrsim 4000$ km/s. 

In the following we limit our calculations to the case of a gas density upstream of the shock $n_{0}=1~\rm cm^{-3}$ and a shock velocity $v_{\rm sh}=2500$ km/s, and we calculate the spectrum of gamma radiation from pp collisions for different values of the neutral fraction $f_{N}$. For a shock velocity $v_{\rm sh}=2500$ km/s, the upstream amplified field turns out to be $\delta B\sim 4\mu G$, and the maximum energy of accelerated particles is $\sim 10-50$ TeV (depending on whether the condition on finite acceleration or finite extent of the upstream is used). In the calculations we use $E_{max}=10$ TeV, since it turns out to describe the preliminary data recently released by VERITAS \citep{VeritasICRC15}. 

It is worth recalling that the maximum energy is mainly affected by the value of the amplified magnetic field upstream of the shock, while depending weakly on the magnetic field downstream. On the other hand, the latter is very important for the multifrequency emission of the remnant. This means that our calculations can only impose a lower limit of $\sim 20\mu$G to the downstream magnetic field. However, additional instabilities, such as the one proposed by \cite{GiacaloneJokipii07}, could further amplify the downstream magnetic field, affecting the synchrotron emission of electrons, but leaving the maximum energy estimated above rather unaffected.

\begin{figure}
  \centering 
  \includegraphics[width=\linewidth]{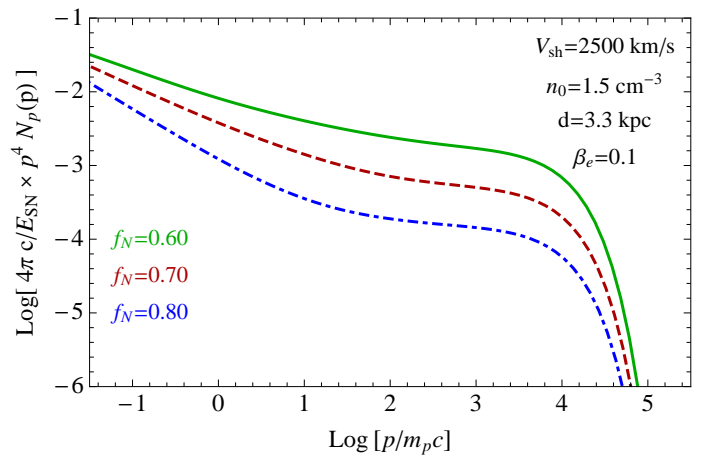}
  \caption{Spectra of protons accelerated at the northeastern limb of the Tycho SNR for a shock velocity $v_{\rm sh}=2500$ km/s and upstream total density $n_{0}=1~\rm cm^{-3}$. The three lines refer to neutral fraction $f_{N}=0.8$ (solid line), $f_{N}=0.7$ (dashed line), and $f_{N}=0.6$ (dash-dotted line). A cutoff in the spectra of accelerated protons at $\sim 10$ TeV is used and reflects the non-resonant magnetic field amplification of magnetic field, as studied by \cite{bell2004}. 
}
  \label{fig:spectra}
\end{figure}

\begin{figure}
  \centering 
  \includegraphics[width=\linewidth]{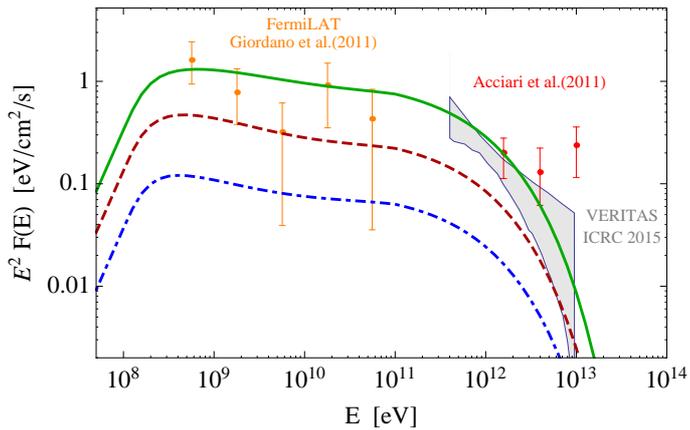}
   \caption{Spectra of gamma rays from production and decays of neutral pions for the proton spectra shown in Fig. \ref{fig:spectra}. The data points show the results of Fermi-LAT \cite[]{fermi} (diamonds) and Veritas \cite[]{veritas} (circles). The shaded area represents the preliminary data recently presented by the Veritas collaboration \citep{VeritasICRC15}.}
  \label{fig:gamma}
\end{figure}

In Fig. \ref{fig:spectra} we show the spectra of accelerated protons for $f_{N}=0.6$ (solid line), $f_{N}=0.7$ (dashed line), and $f_{N}=0.8$ (dash-dotted line). The slightly different normalisation reflects the fact that varying the neutral fraction induces a non-linear reaction that reduces the acceleration efficiency. The spectra are easily seen to be rather steep at energies below a few hundred GeV, while hardening at higher energies: the steeper spectra are the consequence of the NRF but, as discussed above, this effect gradually vanishes when the diffusion length of particles becomes comparable with the ionisation length of neutrals, so that at $E\gtrsim 1$ TeV the spectra show the concavity that is typical of CR modified shocks. Clearly the gamma ray emission reflects this shape, as shown in Fig. \ref{fig:gamma}, where the curves are labelled as in Fig. \ref{fig:spectra}.

Data points are from Fermi-LAT \cite[]{fermi} (diamonds) and VERITAS \cite[]{veritas} (circles). The shaded area represents the preliminary data recently presented by the VERITAS collaboration \citep{VeritasICRC15}. One can see that for reasonable values of the neutral fraction, the spectral slope of accelerated particles is $\sim 2.2$ as required by gamma ray observations but also, maybe in a more compelling manner, by radio observations. The possibility that the spectrum of gamma rays may reflect the presence of neutral hydrogen in the acceleration region relies on the assumption that there may be regions of the Tycho SNR where the shock is appreciably slower than average \cite[possibly as a consequence of higher ISM density as inferred by][]{Williams13}. This also implies that the bulk of the gamma ray emission, at least at energies $E_{\gamma}\lesssim 100$ GeV, is expected to be produced in such a region of slower shock motion in a denser, partially neutral gas. 

To reflect these requirements, we assumed here that about half of the shock is expanding in a denser medium with density $n_{0}=1.5~\rm cm^{-3}$. While a detailed fit to the data would have required also taking the gamma ray emission into account from the other half of the remnant, we decided here to avoid such a complication since it is not our purpose to achieve a quantitative fit to the data. 
Nevertheless, the question remains of whether limiting the calculation of the gamma ray emission to half of the remnant may be consistent with observations from Tycho at other wavelengths. The gamma ray emission from the faster region of the remnant is expected to have a harder and concave shape (if no attempt is made to introduce the velocity of the scattering centres) and a lower normalisation because of the lower gas density. On the other hand, the CR acceleration efficiency might be somewhat higher because of the faster shock motion, although it is unlikely that this efficiency can become higher than $\sim 20\%$ because of the various self-regulation processes that take place during the acceleration process. 

Observations in the spatially integrated radio emission seem to suggest that the electron spectrum in the $\sim$ GeV energy region is somewhat steeper than $E^{-2}$ \cite[]{tychoradio}. However, the morphology of the radio emission shows that the slower NE region of the remnant is about twice as bright as the faster region, thereby suggesting that even the radio emission could be dominated by electrons accelerated in the region where neutral hydrogen is present. This scenario requires that the magnetic field downstream of the NE region of the shock be amplified due to hydrodynamical instabilities (e.g. \cite{GiacaloneJokipii07}). As discussed above, this amplified field would have little impact on the maximum energy (if any, $E_{max}$ would slightly increase) but would increase the radio emissivity and limit the inverse Compton contribution to the observed gamma ray emission). The best way to check the correctness of this ansatz is to measure the spatially resolved spectrum of the radio emission and test the possibility that the radio spectrum is steeper in the NE region. The only attempt to carry out a spatially resolved spectral analysis from Tycho \cite[]{Katz-Stone00} provided results that are not conclusive.

\section{Conclusions}
\label{sec:conclude}

It was first shown by \cite{paper1} that the presence of neutral hydrogen in the acceleration region of SNRs may strongly affect the spectrum of accelerated particles. The effect is the highest for shock velocities $\lesssim 4000$ km/s for which the path length for CE and ionisation are short enough to allow for the deposition of energy and momentum back in the downstream plasma caused by the NRF. For this phenomenon to be effective, these processes must take place on spatial scales that are smaller than the diffusion length of accelerated particles upstream. For typical values of the parameters, this happens for CR energies $\lesssim 1$ TeV. If the NRF deposits energy and momentum upstream, the shock is weakened and the spectrum of accelerated particles steepens. At high enough energies, the standard spectrum of DSA is recovered. The non-linear theory of particle acceleration, where both the NRF and the dynamical reaction of accelerated particles are taken into account, was presented by \cite{paper3}. 

In this paper we applied this theoretical framework to the Tycho SNR. We built upon previous claims that some regions of the Tycho SNR shock may be moving appreciably slower than the average velocity \citep{Reynoso97, Katsuda10}, and we pointed out that the spectrum of accelerated particles in such regions may be steeper than the canonical $E^{-2}$ expected from DSA. This also reflects in steeper gamma ray spectra, which compare well with data from Fermi-LAT \cite[]{fermi} and Veritas \cite[]{veritas}. 

The existence of spectra of accelerated particles that are steeper than expected are rather widely accepted now \cite[]{damiano2011,FermiCat} and a generic explanation of such findings is necessary. Explanations of the steeper spectra range from morphological considerations \cite[see][for a specific application to the Tycho SNR]{berezhkotycho} to speculations about the velocity of scattering centres \cite[]{damiano2012}. Since the presence of neutral hydrogen is quite widespread in the Galaxy, it seems plausible that the steeper spectra of accelerated particles may be due to the neutral induced precursor, as discussed above. This possibility can be tested by using spatially resolved measurements of the radio spectrum, which may show the existence of electrons with harder spectra where the shock is faster (and the NRF is suppressed) and steeper spectra where the shock is slower. For closer SNRs, where spatially resolved gamma-ray observations may be possible, a similar test can be performed in the gamma ray band.

\section*{Acknowledgements}
We are grateful to E. Amato for a careful reading of the manuscript and continuous insightful conversations. We are also thankful to the referee, Don Ellison, for raising several interesting problems with the initial version of the manuscript. This work was partially funded through Grant PRIN-INAF 2012.

\end{document}